\documentclass[twocolumn,showpacs,amsmath,amssymb]{revtex4}
\usepackage{graphicx}
\usepackage{dcolumn}
\usepackage{bm}

 \def\d{\partial} \def\l{\left(} \def\r{\right)}
  \newcommand{\be}{\begin{equation}}
\newcommand{\ee}{\end{equation}} \newcommand{\bea}{\begin{eqnarray}}
\newcommand{\eea}{\end{eqnarray}} \newcommand{\bg}{\begin{gather}}
\newcommand{\eg}{\end{gather}}
\newcommand{\bseq}{\begin{subequations}}
\newcommand{\eseq}{\end{subequations}}  

\renewcommand{\ln}{\mathop{\rm ln}\nolimits}

\begin{document}


\title{Massive graviton as a testable cold dark matter candidate}
\author{S.L.~Dubovsky$^{a,c}$} 
\author{P.G.~Tinyakov$^{b,c}$}
\author{I.I.~Tkachev$^{a,c}$} 
\affiliation{ $^a$Department of Physics, CERN
Theory Division, CH-1211 Geneva 23, Switzerland\\ $^b$Service de Physique
Th\'eorique, Universit\'e Libre de Bruxelles, CP225, bld.~ du Triomphe, B-1050
Bruxelles, Belgium\\ $^c$Institute for Nuclear Research of the Russian Academy
of Sciences, 60th October Anniversary Prospect, 7a, 117312 Moscow, Russia }%
\begin{abstract}
We construct a consistent model of gravity where the tensor graviton
mode is massive, while linearized equations for scalar and vector
metric perturbations are not modified. The Friedmann equation acquires
an extra dark-energy component leading to accelerated expansion. The
mass of the graviton can be as large as $\sim
(10^{15}\mbox{cm})^{-1}$, being constrained by the pulsar timing
measurements. We argue that non-relativistic gravitational waves can
comprise the cold dark matter and may be detected by the future
gravitational wave searches.
\end{abstract}

\maketitle 

{\it 1. Introduction.}  The current cosmological model is in a
beautiful agreement with the data \cite{CMB}. However, it requires
introduction of exotic density components (dark matter, dark energy)
with abundances highly tuned to the baryonic matter. This motivates
interest in modified theories of gravity deviating from the Einstein
theory at large distance scales. Generically, in such theories the
graviton has a non-zero mass. The common lore is that the inverse
graviton masses significantly smaller than the current Hubble scale
are not phenomenologically allowed. In this paper we demonstrate that
the inverse graviton mass can be not only significantly smaller than
the current size of the Universe, but even many orders of magnitude
smaller than the galactic scales.  We argue that massive graviton
provides specific signatures for gravitational wave experiments and
may even account for the cold dark matter (CDM) in the Universe.

Recent studies of the Fierz--Pauli theory of massive
gravity~\cite{Fierz:1939ix} and brane world scenarios where the
four-dimensional graviton has a non-zero
mass~\cite{Gregory:2000jc,Dvali:2000hr} strongly
suggest~\cite{Arkani-Hamed:2002sp,Pilo:2000et,Dubovsky:2002jm,Luty:2003vm,
Rubakov:2003zb,Dubovsky:2003pn,Chacko:2003yp} that Lorentz-invariant
models of massive gravity suffer either from the presence of ghosts
(fields with a wrong sign of the kinetic term), or from the vDVZ
discontinuity due to extra graviton
polarizations~\cite{vanDam:1970vg,Zakharov} and strong coupling at the
low energy scale.  It is possible that the account for the effects of
local curvature may solve these problems in some models
\cite{Vainshtein:1972sx,Deffayet:2001uk,Damour:2002gp,Nicolis:2004qq}. Another
possibility which attracted attention very
recently~\cite{Jacobson:2001yj,Arkani-Hamed:2003uy,Rubakov:2004eb,Gripaios:2004ms,Dubovsky:2004sg}
is to allow for a violation of Lorentz invariance. In particular, a
class of models was found \cite{Dubovsky:2004sg} where tensor graviton
mode is massive, vDVZ discontinuity and strong coupling problems are
absent, while the absence of ghosts and rapid classical instabilities
is ensured by the residual reparametrization symmetry
\begin{equation}
\label{symmetry}
x^i\to x^i+\xi^i(t), 
\end{equation} 
$x^i$ being the spatial coordinates. These models are the focus of the
current paper.

{\em 2. The model.}  In the covariant formalism of
Ref.~\cite{Dubovsky:2004sg} (see also
Refs.~\cite{Arkani-Hamed:2002sp,Arkani-Hamed:2003uy}), the action for
the theory of massive gravity contains the metric $g_{\mu\nu}$ and four scalar
Goldstone fields $\phi^0,\;\phi^i$ ($i=1,\dots,3$). In the presence of
the residual symmetry (\ref{symmetry}) it reads
\begin{equation}
\label{action}
S=\int d^4x\sqrt{-g}\left[ - M_{Pl}^2
R+\Lambda^4F(X,W^{ij},\dots)\right] ,
\end{equation}
where $X$ and $W^{ij}$ are the scalar quantities constructed from the
Goldstone fields and the metric,
\[
X=g^{\mu\nu}\d_\mu\phi^0\d_\nu\phi^0,\nonumber\\
\]
\begin{equation}
W^{ij}={g^{\mu\nu}\d_\mu\phi^i\d_\nu\phi^j}-
{g^{\mu\nu}\d_\mu\phi^0\d_\nu\phi^i\cdot
g^{\lambda\rho}\d_\lambda\phi^0\d_\rho\phi^j\over X},
\label{XVYW}
\end{equation}
and $F$ is a function to be constrained later.  We assume that the
Goldstone sector is characterized by a single energy scale $\Lambda$.
Dots in Eq.~(\ref{action}) stand for higher-derivative terms.  Latin
indices $i,j$ are contracted using $\delta_{ij}$.

We require that the model admits a background solution with the metric
$g_{\mu\nu}$ equal to the Minkowski metric $\eta_{\mu\nu}$ and the
scalar fields taking the form
\begin{equation}
\label{scalarvevs}
\phi^0=a\Lambda^2 t\;,\;\;\phi^i=b\Lambda^2 x^i 
\end{equation} 
for some constants $a$ and $b$.  For a generic function $F$ such a
solution always exists. In the ``unitary gauge'' where the Goldstone
fields are fixed to their vacuum values (\ref{scalarvevs}), the second
term in the action (\ref{action}) gives rise to the following mass
term for the metric perturbation $h_{\mu\nu}$,
\begin{equation}
\label{LVmass}
{\cal{L}}_m={M_{Pl}^2\over 2}\l m_0^2h_{00}^2-m_2^2h_{ij}^2
+m_3^2h_{ii}^2-2m_4^2h_{00}h_{ii}\r,
\end{equation}
where the values of the mass parameters $m_a$ are determined by the
first and the second derivatives of the function $F(X,W^{ij})$ at the
vacuum values of its arguments as defined by eqs.~(\ref{XVYW}) and
(\ref{scalarvevs}). The overall scale $m$ of the graviton masses is
related to $\Lambda$ as $m \sim \Lambda^2/M_{Pl}$.  The analysis of
Ref.~\cite{Dubovsky:2004sg} implies that $\Lambda$ plays the role of
the cutoff scale of the theory with the action (\ref{action}).

The residual reparametrization symmetry (\ref{symmetry}) arises
in the unitary gauge as a consequence of the global symmetry
$
\phi^i\to\phi^i+ \xi^i(\phi^0)
$
of the covariant action (\ref{action}). This symmetry implies, in
particular, that there is no graviton mass term proportional to
$h_{0i}^2$.

As is usual in the linearized theory, it is convenient to consider
separately tensor, vector and scalar metric perturbations (cf.
Refs.~\cite{Rubakov:2004eb,Dubovsky:2004sg}). The tensor modes ---
transverse traceless gravitational waves $h_{ij}^{TT}$ --- have
non-zero mass equal to $m_2$~\cite{Rubakov:2004eb}. There are no
propagating degrees of freedom in the vector
sector~\cite{Dubovsky:2004sg}.  Moreover, the contribution of the mass
term (\ref{LVmass}) in the vector sector has the form of a gauge
fixing.  Consequently, no modification of gravity arises in the vector
sector at the order we are working.  Finally, the energy-momentum
tensor $\delta T_{\mu\nu}$ induces the following perturbations in the
scalar sector,
\begin{gather}
\label{Psi}
\Psi=\Psi_E,\\
\label{Phi}
\Phi=\Phi_E+{m_2^2[ 3m_4^4-m_0^2\l 3m_3^2-m_2^2\r]\over
m_4^4-m_0^2(m_3^2-m_2^2) }{1\over \d_i^4}
{\delta T_{00}\over M_{Pl}^2},
\end{gather}
where $\Psi$, $\Phi$ are the gauge-invariant scalar potentials defined
in a standard way \cite{Mukhanov:1990me}, and $\Psi_E$, $\Phi_E$ are
their values in the Einstein theory.  The modification of gravity
manifests itself in the last term in Eq.~(\ref{Phi}). There is no vDVZ
discontinuity as this term vanishes in the limit when all graviton
masses uniformly go to zero.

The extra term in Eq.~(\ref{Phi}) grows linearly with the distance
from the source, indicating the breakdown of the linearized
theory. This growth cannot be eliminated by a proper choice of the
gauge as $\Phi$ is the gauge-invariant quantity. However, the Riemann
curvature associated with the extra term goes to zero as $1/r$ at
large $r$, so the space-time becomes flat far from the source. (This
breakdown of perturbation theory is very different in nature from the
seemingly similar problem in the Fierz--Pauli
theory~\cite{Vainshtein:1972sx}, where it happens in the vicinity of
the source. The close analogue of the phenomenon discussed here is the
breakdown of perturbation theory far from the source in the
three-dimensional classical Yang--Mills theory.)  In the region where
the non-standard term in Eq.~(\ref{Phi}) is still small it produces
the $r$-independent force, imitating the effect of a halo with the
density profile $\propto r^{-1}$.

The analysis of Eq.~(\ref{Phi}) in the region where it
enters the non-linear regime goes beyond the scope of this paper.
Instead, we chose the masses $m_a$ in such a way that the second term
in Eq.~(\ref{Phi}) vanishes.  It is important that this can be
achieved by imposing, in addition to (\ref{symmetry}), the following
dilatation symmetry,
\begin{equation}
\label{dilatation}
t\to\lambda t\;,\;\;x^i\to\lambda^{-\gamma} x^i,
\end{equation}
where $\gamma$ is a real  constant.  At the linearized level this
symmetry implies the following relations among masses,
\begin{equation}
\label{dilamasses}
m_0^2=-3\gamma m_4^2\;,\;\; m_2^2-3m_3^2=\gamma^{-1}m_4^2,
\end{equation}
which lead to the cancellation of the second term in Eq.~(\ref{Phi})
for any $\gamma$. Thus, when the symmetry (\ref{dilatation}) is
imposed, the only modification of gravity at the linearized level is
the non-zero mass of the graviton.

The inclusion of higher-derivative terms in the action (\ref{action})
leads in general to the appearance of the dynamical degree of freedom
in the scalar sector~\cite{Dubovsky:2004sg}. This degree of freedom is
similar to that present in the ghost condensate
model~\cite{Arkani-Hamed:2003uy}. It has a healthy kinetic term
provided the following inequality holds~\cite{Dubovsky:2004sg},
\begin{equation}
\label{1deriv}
m_0^2-{m_4^4\over (m_3^2-m_2^2)}>0.
\end{equation}
The latter condition is compatible with Eqs.~(\ref{dilamasses}) and the
requirement that the graviton mass is not tachyonic, $m_2^2>0$.  The
effects related to this degree of freedom are characterized by the
huge retardation time $\sim m^{-1}
(M_{Pl}/\Lambda)$~\cite{Arkani-Hamed:2003uy,Dubovsky:2004qe,Peloso:2004ut}.
This time is larger than the current age of the Universe for the
values of the graviton mass $m$ specified below, so we can consistently
neglect these effects.

In the covariant formalism the residual symmetry (\ref{dilatation})
translates into the following global symmetry of the Goldstone sector,
$
\phi^0\to\lambda \phi^0\;,\;\;\phi^i\to\lambda^{-\gamma} \phi^i.
$
The action invariant under the symmetries (\ref{symmetry}),
(\ref{dilatation}) has the form (\ref{action}) with the function $F$
depending on the single combination $X^\gamma W^{ij}$.
The case of the ghost condensate~\cite{Arkani-Hamed:2003uy} 
emerges in the limit $\gamma \rightarrow \infty$
and requires a fine-tuning of $F$ to obtain the Minkowski vacuum.  
The Minkowski vacuum with the scalar vev's of
the form (\ref{scalarvevs}) exists for a general function $F$ if
$\gamma=1/d$, where $d=3$ is a number of spatial dimensions.  
For definiteness, in what follows we consider the case $F = F(X^{1/3} W^{ij})$.

{\it 3. Cosmological solutions.}  The spatially flat homogeneous
cosmological ansatz is
\begin{gather}
ds^2=a^2(\eta)\l d\eta^2-dx_i^2\r,\\
\phi^0=\phi(\eta)\;,\;\;
\phi^i=\Lambda^2x^i.
\end{gather}
In what follows we assume that the rate of the expansion is much
smaller than the energy scale $\Lambda$, so one can neglect higher
derivative terms in the action (\ref{action}). For simplicity, let us
also assume that the function $F$ depends only on the combination
$Z\equiv X^{1/3}W^{ij}\delta_{ij}$.
The Einstein equations are reduced to the Friedmann equation
\begin{gather}
\label{00eq}
\l{\dot{a}\over a^2}\r^2={1\over 3M_{Pl}^2}\l\rho_m+{2\over 3}\Lambda^4
F'(Z)Z-\Lambda^4F(Z)\r,
\end{gather}
where $\rho_m$ is the energy density of matter,
and the field equation for $\phi^0$,
\begin{equation}
\label{phi0}
\d_\eta\l {a^3F'(Z)W X^{-1/6}}\r=0.
\end{equation}
Eq.~(\ref{phi0}) implies $Z=const$ or, equivalently, $
\phi^0\propto\int d\eta a^4(\eta)$. Then Eq.~(\ref{00eq}) takes the
form of the standard Friedmann equation with the value of the
cosmological constant determined by the value of $Z$, {\it i.e.}  by
the initial conditions in the Goldstone sector. Note that these
initial conditions may be different in different regions of
space. Therefore, this model is an example of the setup where de
Sitter solutions with different expansion rates exist for any value of
the vacuum energy. This property is a welcome feature for the
application of the weak anthropic principle \cite{Weinberg:1988cp} to
the cosmological constant problem.

To summarize, we have constructed a consistent model where
gravitational waves are massive, while linearized equations for the
metric perturbations in the scalar and vector sectors, as well as
spatially flat cosmological solutions, are the same as in the Einstein
theory. In this model, the tests of (linear) gravity based on the
solar system and Cavendish-type experiments \cite{Esposito-Farese:2000ij} are
automatically satisfied, while the main constraints are coming from
emission and/or propagation of gravitational waves.

{\it 4. Relic gravitational waves.}  Observations of the slow down of
the orbital motion in binary pulsar systems~\cite{pulsars} imply that
the mass of the gravitational waves cannot be larger than the
frequency of the waves emitted by these systems.  The latter is
determined by the period of the orbital motion which is of order 10
hours, implying the following limit on the graviton mass,
\begin{equation}
\label{pulsarlimit}
{m_2\over 2\pi}\equiv \nu_2\lesssim  3\cdot 10^{-5}~{\mbox{Hz}}\approx
\l 10^{15}~{\mbox{cm}}\r^{-1}.
\end{equation}

Let us estimate the cosmological abundance of relic gravitons.  For
this purpose we consider the transverse traceless perturbation of the
metric $h_{ij}$.  The quadratic action for $h_{ij}$ in the expanding
Universe takes the following form,
\begin{equation}
M_{Pl}^2\int d^3kd\eta a^2(\eta)
\l 
\dot h^{2}_{ij}-
\l\d_kh_{ij}\r^2-m_2^2a^2(\eta)h^2_{ij}\r.
\end{equation}
This has a form of the action for a minimally coupled massive scalar
field. Therefore, gravitons in our model are produced efficiently
during inflation (cf. Ref.~\cite{Rubakov:1982df}).

To be concrete, consider a scenario where the
Hubble parameter $H_i$ is constant during inflation.
This scenario may be realized, for instance, in hybrid models of
inflation~\cite{Linde:1993cn}.  First, we need to check that the
phenomenologically relevant values of parameters correspond to the
regime below the cutoff scale of the effective theory,
i.e. $H_i\lesssim\Lambda$.  For the energy scale of inflation
$E_i\sim\sqrt{H_iM_{Pl}}$ this implies
\begin{equation}
\label{energy}
E_i<m_2^{1/4}M_{Pl}^{3/4}\approx 10^7~{\mbox{GeV}}
\l m_2\cdot 10^{15}~{\mbox{cm}}\r^{1/4}.
\end{equation}
This value is high enough to allow for a successful baryogenesis even
for graviton masses of the order of the current Hubble scale.

Consider now the production of massive gravitons.  Assuming the above
scenario of inflation, the perturbation spectrum for the massive
gravitons is that for the minimally coupled massive scalar field in
the de Sitter space~\cite{Bunch:1978yq},
\begin{equation}
\label{spectrum}
\langle h_{ij}^2\rangle\simeq{1\over 4\pi^2} \l {H_i\over M_{Pl}}\r^2\int
 {dk\over k}
\l{k\over H_i}\r^{2m_2^2\over 3H^2}\;.
\end{equation}
Superhorizon metric fluctuations remain frozen until the Hubble factor becomes
smaller than the graviton mass, when they start to
oscillate with the amplitude decreasing as $a^{-3/2}$.  The energy
density in massive gravitons at the beginning of oscillations is of
order
\begin{equation}
\rho_o\sim M_{Pl}^2m_2^2\langle h_{ij}^2\rangle\simeq{3 H_i^4\over 8\pi^2}\;,
\end{equation}
where we integrated in Eq.~(\ref{spectrum}) over the modes longer than the
 horizon.
Today the fraction of the energy
density in the massive gravitational waves is 
\begin{equation}
\label{Omega}
\Omega_g={\rho_o\over z_o^3\rho_c}=
{\rho_o\over z_e^3\rho_c}\l H_e\over H_o\r^{3/2},
\end{equation}
where $z_o$ is the redshift at the start of oscillations,
$H_o\sim m_2$ is the Hubble parameter at that time, 
$H_e\approx 0.4\cdot 10^{-12}$~s$^{-1}$
 is the Hubble parameter at the matter/radiation equality, and $z_e\approx 
3200$
is the corresponding redshift.
Combining all the factors together one gets
\begin{equation}
\label{finalOmega}
\Omega_g\sim 3\cdot 10^3 (m_2\cdot 10^{15}{\mbox{cm}})^{1/2}\l H_i\over
\Lambda\r^4.
\end{equation}
This estimate assumes that the number of e-foldings during inflation
is large, $\ln N_e>H^2/m^2$, which is quite natural in the
model of inflation considered here. 

According to Eq.~(\ref{finalOmega}), the massive gravitons are produced
efficiently enough to comprise all of the cold dark matter, provided the value
of the Hubble parameter during inflation is about one order of magnitude below
the scale $\Lambda$.  We find it encouraging that one obtains $\Omega_g\sim 1$
when the initial energy density in the metric perturbations is close to the
cutoff scale, $\rho_o^{1/4}\sim\Lambda$. This suggests that other mechanisms
of production unrelated to inflation (e.g., similar to those invoked for the
axion or Polony fields) may naturally lead to the same result, $\Omega_g\sim
1$.

The produced gravitons may cluster in galaxies. To account for the
dark matter in galactic halos the graviton mass should satisfy $
(mv)^{-1}\lesssim 1~\mbox{kpc}\sim 3\cdot 10^{21}~\mbox{cm} $, where
$v\sim 10^{-3}$ is a typical velocity in the halo.

{\it 5. Detection.}  Let us now briefly describe potential
observational signatures of the above scenario. Note first that at
distances shorter than the wavelength, the effect of a transverse
traceless gravitational wave on test massive particles in Newtonian
approximation is described by the acceleration $\ddot{h}_{ij}x^j/2$
(see, e.g., Ref.~\cite{Thorne:1987af} for a review). The same is true
for massive gravitational waves, the only difference being that the
wavelengths are longer in the non-relativistic case, so the Newtonian
description works for the larger range of distances. Thus, the
non-relativistic waves act on the detector in the same way as massless
waves of the same frequency.

Let us estimate the amplitude of the gravitational waves assuming that they
comprise all of the dark matter in the halo of our Galaxy.
The energy density in non-relativistic gravitational waves is of order
$M_{Pl}^2m_2^2h_{ij}^2$. 
Equating this to the local halo density
one gets
\begin{equation}
\label{hestimate}
\langle
h_{ij}\rangle
\sim 10^{-10}\l{3\cdot 10^{-5}\mbox{Hz}
\over  \nu_2}\r.
\end{equation}
At the frequencies $ 10^{-6}\div 10^{-5}$~Hz this value is well above
the expected sensitivity of the LISA detector~\cite{Bender:2003uv}.
Note that in the close frequency range $10^{-9}\div 10^{-7}$~Hz there
is a restrictive bound~\cite{timing} at the level $\Omega_g<10^{-9}$
on the stochastic background of the gravitational waves coming from
the timing of the millisecond pulsars~\cite{sazhin}. So, it is
possible that our scenario can be tested by the reanalysis of the
already existing data on the pulsar timing.

The relic abundance of gravitons may depend on both the specific
inflationary model and the details of the (unknown) UV completion of
massive gravity. In general, massive gravitons may not comprise the
whole of the CDM in the galaxy halos.  It is important that the
expected LISA sensitivity allows to detect the presence of massive
gravitons at the significantly lower level than in
Eq.~(\ref{hestimate}).

{\it 6. Concluding remarks.}  In this paper we limited ourselves to a
specific choice of the parameters (graviton masses and the constant
$\gamma$ entering Eq.~(\ref{dilatation})) such that there is no
modification of the Newton potential at the linear level, and the
cosmological evolution remains standard. We also did not consider
possible non-linear effects, which may become a necessity at different
choice of the parameters.  A number of interesting questions is
related to these effects, including the limits on graviton masses,
clustering of massive gravitons in haloes and proper modifications of
Eqs.~(\ref{00eq}) and (\ref{phi0}) to account for the direct coupling
between Goldstone fields and gravitons. We expect, however, that our
main conclusions --- that gravitons may have large masses and may be
produced with cosmologically significant abundance --- are generic in
this class of models. In the relevant range of parameters, a specific
signature of the gravitons with non-zero mass is a strong
monochromatic signal in the detectors of gravitational waves.  An
independent measurement of the graviton mass may be performed at
future gravitational wave detectors (for a review, see,
e.g.~\cite{Maggiore:1999vm}) operating at higher frequencies by
testing the delay between the electromagnetic and gravitational
signals from a distant supernova explosion.

{\it Acknowledgments.} 
We thank M.~Maggiore,
R.~Rattazzi, V.~Rubakov,  M.~Sazhin, M.~Shaposhnikov 
for useful discussions. SD thanks ULB (Bruxelles) and SPhT (Saclay) where
part of this work was done for a warm hospitality. The work of P.T. is
supported by IISN, Belgian Science Policy (under contract IAP V/27).

\end{document}